\documentclass[a4paper]{PoS}

\usepackage{gensymb}
\usepackage{booktabs}
\usepackage[caption=false]{subfig}
\usepackage{graphicx}

\title{The ultra-high energy cosmic rays image of Virgo A}

\ShortTitle{The UHECRs image of Virgo A}

\author{\speaker{Radom\'{\i}r \v{S}m\'{\i}da}\\
        Karlsruhe Institute of Technology, Germany\\
        E-mail: \email{radomir.smida@kit.edu}}

\author{Ralph Engel\\
        Karlsruhe Institute of Technology, Germany\\
        }

\abstract{Arrival directions of ultra-high energy cosmic rays from the direction
of ten brightest radio sources lying within $50$~Mpc from our Galaxy were studied
by using recent models of the large-scale Galactic magnetic field. A detailed
study, where also small-scale turbulent magnetic field component was implemented,
is presented for the radiogalaxy Virgo~A. This radiogalaxy is located far from
the Galactic plane which leads to a unique image of this UHECR source candidate,
if the flux is composed from a mixture of intermediate mass nuclei. We present
a method suitable for identifying cosmic rays arriving from this close-by
radiogalaxy.
}

\FullConference{The 34th International Cosmic Ray Conference,\\
		30 July- 6 August, 2015\\
		The Hague, The Netherlands}

\begin{document}

\section{Introduction}

The search of a point source of ultra-high energy cosmic rays (UHECRs),
defined as cosmic rays with energy above $50$\,EeV in this paper,
lacks an answer more than half-century after their discovery~\cite{Linsley1961,Linsley1963}.
The primary particles are charged nuclei accelerated in astronomical objects.
Moreover, the distance to such sources must be within $\sim100$\,Mpc for UHECRs
because of energy-loss processes on various photon backgrounds~\cite{Greisen1966,Zatsepin1966}.

Categories of astronomical objects powerful enough to accelerate nuclei up
to the highest measured energy~\cite{Bird1995} have been classified in the
past~\cite{Hillas1984}.
For the search of the site of the UHECR origin an identification of a local point
source is of the primary concern, because the flux from such source
located within a few tens of Mpc won't be significantly attenuated for most
of nuclei~\cite{Kotera2011,Allard2012}. Moreover, extragalactic magnetic fields
might be rather uniform over such distance. The most interesting position of
the source in the sky is at high Galactic latitude, because the large-scale
Galactic magnetic field (GMF) between the source and the Earth has
the least complex structure in this direction.

Any nuclei of an electric charge $Z$ will be deflected during their propagation
through the GMF. Under an assumption that all particles from the same source
travel through the same magnetic field the angular deflection will scale linearly
with the electric charge. It is therefore useful to study the angular deflection
as a function of the magnetic rigidity $R(\mathrm{V})=E(\mathrm{eV})/Z$ instead
of energy $E(\mathrm{eV})$.

We will assume that UHECRs arrive to the Galaxy in a parallel beam from an
extragalactic point source and information about the source position is not
lost during the propagation in the extragalactic space. Under such assumptions
we can study an effect of the large-scale GMF and also small-scale turbulent
magnetic fields on the source image at the Earth.

\section{Simulations}

Point source candidates and the GMF model must be selected before performing
a simulation. For the former the full-sky the catalogue of radiogalaxies
of the local universe~\cite{Velzen} was selected. This catalogue was prepared
for any study of UHECRs and provides information required for our analysis.
Ten brightest radiogalaxies within the distance $D=50$~Mpc have been selected
from this catalogue and their list is provided in Tab.~\ref{Tab-radiogalaxies}.
The closest object from our selection is Cen~A at the distance $3.6$~Mpc followed
by Virgo~A and Fornax~A.

These ten celestial objects are not uniformly distributed across the sky,
because they follow the local mass distribution. For example the sky positions
of radiogalaxies NGC~5090 and Cen~A are separated by only $1\degree$ and
in the case of Virgo~A and NGC~4261 only $7\degree$.

\begin{table}
\begin{center}
\begin{tabular}{c c c c c c}
\toprule
Name & $\alpha$ & $\delta$ & $l$ & $b$ & $D$ \\
\midrule
Fornax A & 50.7 & -37.2 & 240.2 & -56.7 & 20.90\\
Virgo A & 187.7 & 12.4 & 283.8 & 74.5 & 18.44\\
NGC 5090 &  200.3 & -43.7 & 308.6 & 18.8 & 46.89\\
Cen A & 201.4 & -43.0 &  309.5 & 19.4 & 3.59\\
NGC 4261 & 184.8 & 5.8 & 281.8 & 67.4 & 32.15\\
NGC 4696 & 192.2 & -41.3 & 302.4 & 21.6 & 41.71\\
IC 5063 & 313.0 & -57.1 & 340.0 & -38.7 & 46.63\\
NGC 5793 & 224.9 &  -16.7 & 342.0 & 36.3 & 46.86\\
NGC 2663 & 131.3 & -33.8 & 255.7 & 5.6 & 29.97\\
NGC 7626 & 350.2 & 8.2 & 87.9 & -48.4 & 47.77\\
\bottomrule
\end{tabular}
\caption{
Ten brightest radiogalaxies within $D<50$~Mpc~\cite{Velzen}.
Columns: name of radiogalaxy, right ascension, declination, galactic longitude
and latitude, distance in Mpc.
}
\label{Tab-radiogalaxies}
\end{center}
\end{table}

All simulated particles were randomly generated in a flat circular area
having its centre $50$~kpc from the Galactic centre. The position of the
centre was at the sky position of a radiogalaxy. The area was always
perpendicular to the direction connecting the Galactic centre and the
radiogalaxy. The radius of this area was $20$~kpc and $10$~kpc for our
simulations without and with the turbulent magnetic field, respectively.

The radius of a detector at the position of the Earth was $100$~pc.
The maximum time to track particle was one million years and the minimum
time step was $10$ years. The CRT
uses adaptive Runge-Kutta integration methods to determine the trajectory
of a charged particle through a magnetic field according to the relativistic
Lorentz force. Detailed description of the CRT code can be found in~\cite{CRT}
and in the manual.

\section{Regular GMF}

The Janson\&Farrar (JF12) GMF model~\cite{JF2012} was primarily used, but
also other models were studied~\cite{Sun,Pshirkov}. We will present only
results obtained for the JF12 model in this paper. A parallel beam of particles
pointed towards the Earth was generated in a circular area of $20$~kpc diameter
at the Galactocentric distance $50$~kpc. The studied magnetic rigidity was
between $1$ and $150$~EV and the spectral index was $1$.

\begin{figure}[t!]
\begin{center}
\includegraphics[width=0.6\columnwidth]{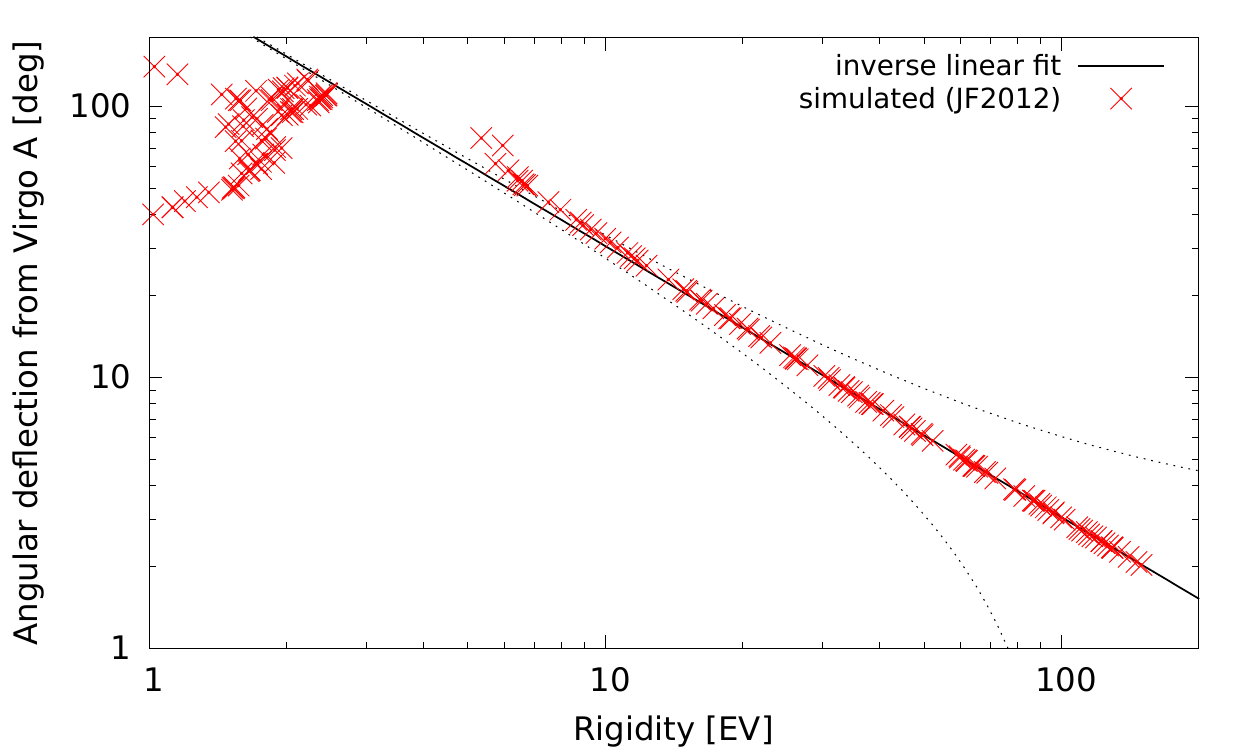}
\caption{Red crosses show the angular deflection from Virgo~A as function
of the rigidity, black solid line shows the inverse linear fit and dashed
lines the $\pm3\degree$ offset from the fit. Notice missing successful
hits between $2.5$ and $5.4$~EV, i.e. when particles cross the Galactic plane.
}
\label{Fig-deflection-m87}
\end{center}
\end{figure}

In a uniform magnetic field the angular deflection is inversely proportional
to the magnetic rigidity $\theta=K/R$. Results of our simulations above $30$~EV
were used to obtain the constant $K$.
Then all results were compared with the inverse linear fit, see for example
Fig.~\ref{Fig-deflection-m87} for Virgo~A.

Our results are summarised in Tab.~\ref{Tab-linear-fit}, where
the numerical constant $K$ and the rigidity where simulated arrival direction
differs more than $3\degree$ from the fit are provided. The uncertainty of values
in Tab.~\ref{Tab-linear-fit} is $\pm5\degree$~EV for the constant $K$ and
$\pm1$~EV for $R_{3\degree}$. Two values of $R_{3\degree}$ are provided
for NGC~2663, because of two different trajectories hitting the Earth
at the same rigidity. The same happens for NGC 4696 and NGC 5090,
but only below $30$~EV. (Let us notice, that two possible trajectories
at the same rigidity have been found even for Virgo A and other GMF
model, namely Sun et al.~\cite{Sun}.)

\begin{table}
\begin{center}
\begin{tabular}{c c c c c c}
\toprule
Radiogalaxy & $K$~($\degree$~EV) & $R_{3\degree}$~(EV) \\
\midrule
Fornax A & 242 & 4 \\
Virgo A & 306 & 7 \\
NGC 5090 & 183 & 16 \\
Cen A & 191 & 15 \\
NGC 4261 & 274 & 8 \\
NGC 4696 & 149 & 12 \\
IC 5063 & 330 & 10 \\
NGC 5793 & 370 & 14 \\
NGC 2663 & 192 & 10, 25 \\
NGC 7626 & 413 & 10 \\
\bottomrule
\end{tabular}
\caption{
Results for the JF12 model of the regular GMF. Parameters of the linear
fit described in the text are provided.
}
\label{Tab-linear-fit}
\end{center}
\end{table}

Results obtained for the JF12 model, see Tab.~\ref{Tab-linear-fit}, show
that the angular deflection is $2\degree-4\degree$ at $R=100$~EV for
all studied radiogalaxies. The simulated data are described by the inverse
linear fit below $10$~EV only for radiogalaxies lying far from the Galactic
plane, i.e. $\vert b \vert>50\degree$. All other radiogalaxies have
$R_{3\degree}\geq10$~EV, i.e. the energy of $70$ and $140$~EeV for nitrogen
and silicon nuclei, respectively.

\begin{figure}[t!]
\begin{center}
\includegraphics[width=0.6\columnwidth]{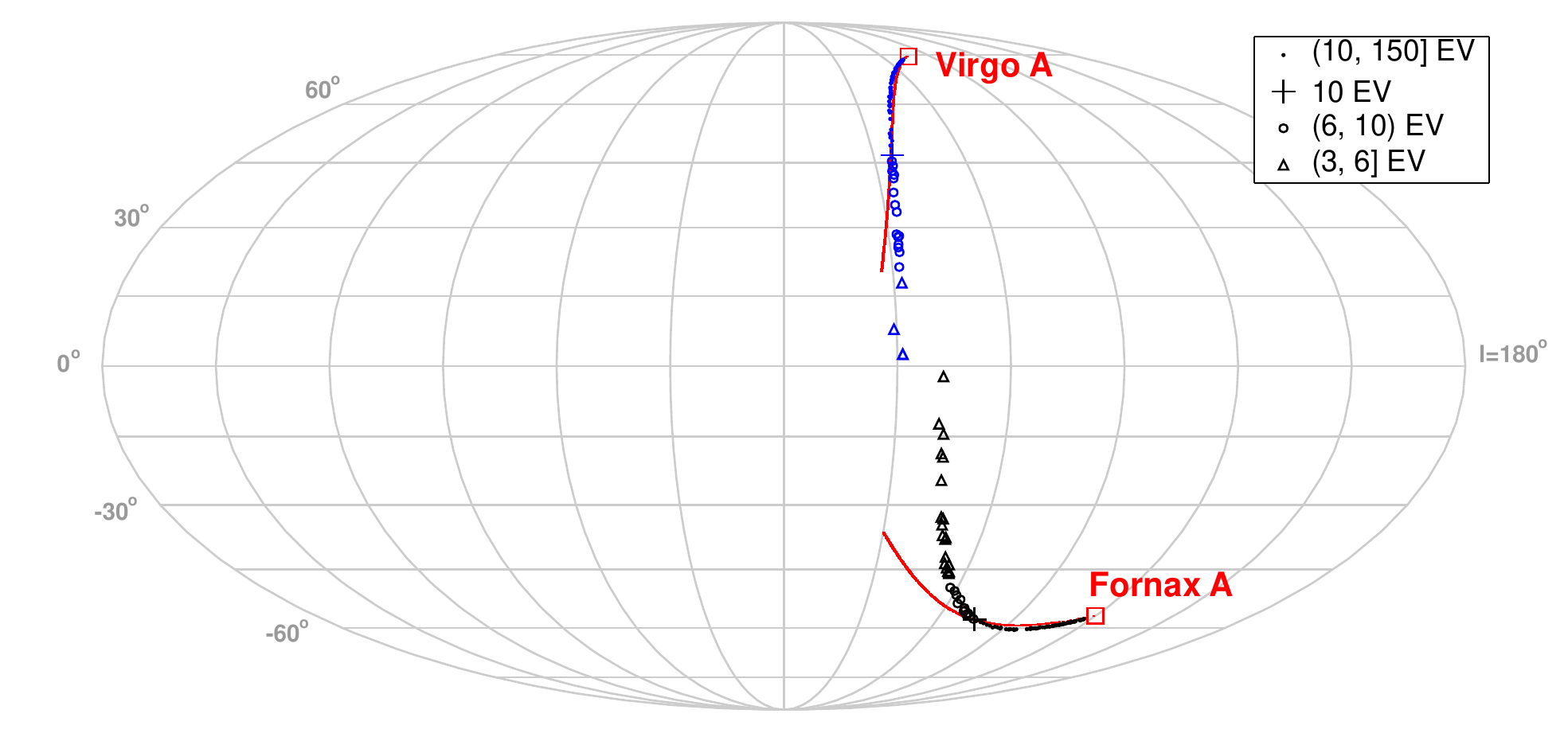}
\caption{
Arrival directions of cosmic rays from Virgo~A and Fornax~A, indicated by
big red squares. The Galactic coordinates are used and the Galactic centre
is in the centre of this map. Blue (black) colour is used Virgo~A (Fornax~A).
For clarity only events above $3$~EV are shown. 
}
\label{Fig-Map-m87-fornaxa}
\end{center}
\end{figure}

Interestingly, the lowest value of $R_{3\degree}$ has the strongest
local radiogalaxy Fornax~A. However, by checking the sky map one can
recognise that even if the angular deflection scales with $R$ the direction
abruptly turns at $R\simeq10$~EV, see Fig.~\ref{Fig-Map-m87-fornaxa}.

It is reasonable to study a location of a point-source only at rigidities,
where the information is not completely lost or vastly reduced by the propagation.
We can identify this rigidity with $R_{3\degree}$.

\begin{figure}[t!]
\begin{center}
\includegraphics[width=0.6\columnwidth]{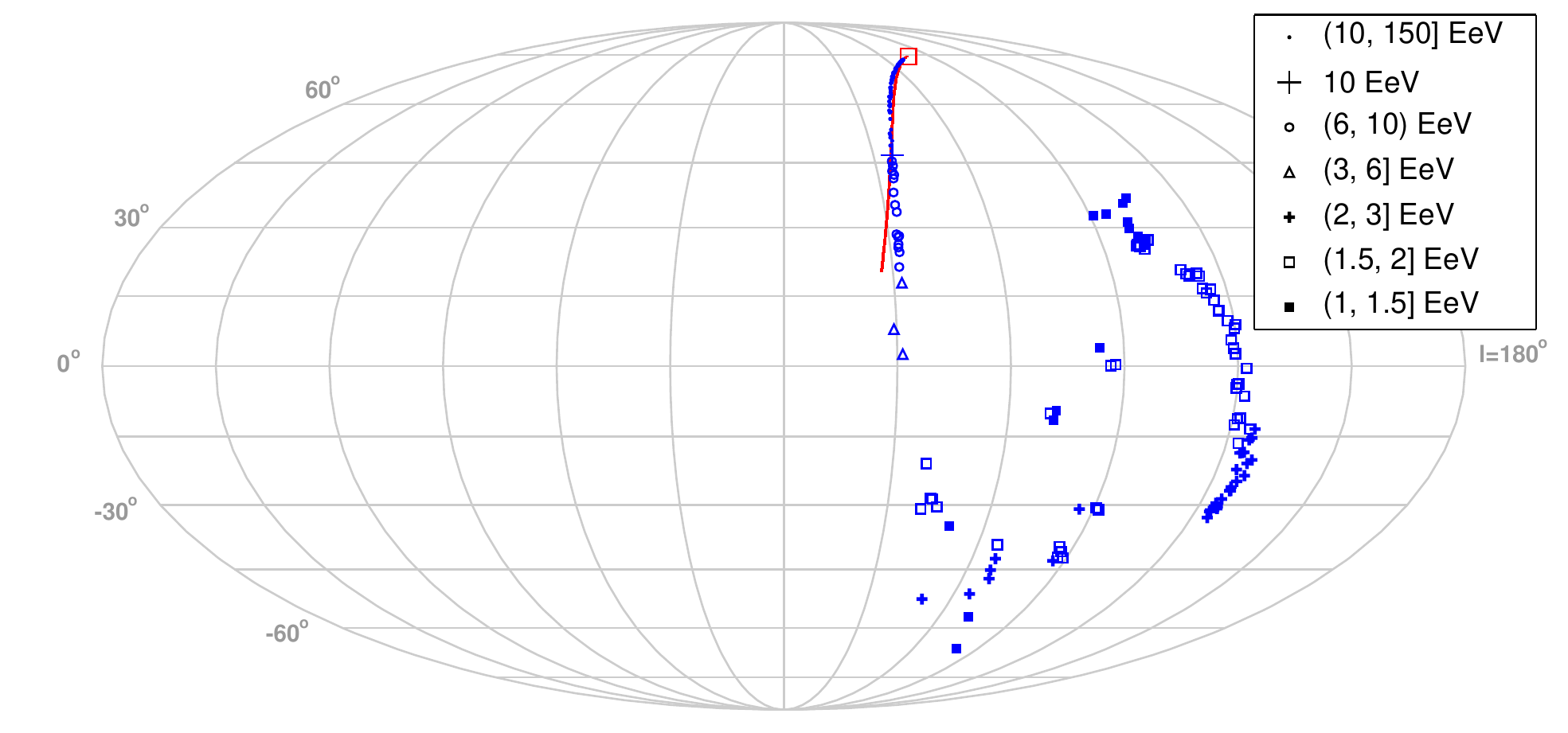}
\caption{
Arrival directions of cosmic rays from Virgo~A between $1$ and $150$~EV.
Red line connects sky positions of a radiogalaxy and $10$~EV event and its
length is $60\degree$.
}
\label{Fig-Map-m87}
\end{center}
\end{figure}

In the case of Virgo~A the simulated results above $7$\,EV agree within
$3\degree$ with values obtained from the inverse linear fit as can be seen
in Fig.~\ref{Fig-deflection-m87} and Tab.~\ref{Tab-linear-fit}.
An increase of discrepancy between calculated and simulated values below
$R_{3\degree}$ is caused by more complex structure of the GMF in the vicinity
of the Galactic plane. Due to this plane none arrival of simulated particles
between $2.5$ and $5.4$\,EV was registered at the Earth and arrival directions
of particles at rigidities below $2.5$~EV are highly scattered, see Fig.~\ref{Fig-Map-m87}.

\section{Turbulent magnetic field}

From ten studied local radiogalaxies the arrival deflection follows a single
arc and the deflection can be described by the inverse linear fit below $R=10$~EV
only for two of them, Virgo~A and NGC~4261. For further investigation of the UHECR
image we have selected the radiogalaxy Virgo~A.

The small-scale random magnetic fields have to be included to the large-scale
regular GMF to obtain more realistic description of arrival directions.
We used four realisations of random magnetic field characterised by
a Kolmogorov spectrum and generated in CRT. Four fields of spheres with
sizes and field magnitude drawn from Gaussian distributions and each
a randomly oriented field vector were generated. Each random magnetic
field realisation is described by eight parameters: number of divisions
within the KRF box in each space coordinate axis ($NX$, $NY$, $NZ$),
length of each side of the KRF box ($LX$, $LY$, $LZ$), the minimum and
maximum wavelengths ($LMI$ and $LMX$). Values used in our simulations
are in Tab.~\ref{Tab-krf}.

\begin{table}
\begin{center}
\begin{tabular}{c c c c}
\toprule
Realisation & $NX,NY,NZ$ & $LX,LY,LZ$~(kpc) & $LMI,LMX$~(pc) \\
\midrule
KRF1 & $256,256,256$ & $5.12,5.12,5.12$ & $5, 100$\\
KRF2 & $256,256,256$ & $5.12,5.12,5.12$ & $5, 100$\\
KRF3 & $256,256,256$ & $5.12,5.12,5.12$ & $5, 512$\\
KRF4 & $256,256,256$ & $2.56,2.56,2.56$ & $5, 100$\\
\bottomrule
\end{tabular}
\caption{
Eight parameters describing four magnetic field realisations. Two
realisations (KRF1 and KRF2) have the same parameters, but different
seed number was used for them.
}
\label{Tab-krf}
\end{center}
\end{table}

As for the regular GMF study a parallel beam of particles was generated
in a circle centred at the sky position of Virgo~A at the galactocentric
distance of $50$~kpc. The differences from the previous case were as follows:
the diameter was $10$~kpc, the rigidity was $5-150$~EV and the spectral
index was $3$. The lower limit of the rigidity was set at $5$~EV to avoid
trajectories crossing the Galactic plane. Simulations were stopped when at
least $90$ hits were reached.

The least difference between a KRF realisation and the regular JF12 field
was for KRF4, which had the smallest side length among used KRF realisations.
Otherwise, all KRF realisations show similar features: small offset at
the highest rigidities, departure more than $\pm3\degree$ from the inverse
linear fit between $10$ and $20$~EV and large scattering of angular deflections
between $5$ and $8$~EV. All these effects could be expected and they indicate
an increasing importance of the turbulent component of the GMF
on the propagation at rigidities below $\sim10$~EV.

Our results for Virgo~A can be compared with similar study made for Cen~A~\cite{Keivani2014},
which lies close to the Galactic plane ($b=19.4\degree$).

\section{Triangular area}

\begin{figure}[t!]
\begin{center}
\subfloat[JF12 + KRF1]{
\includegraphics[width=0.48\columnwidth]{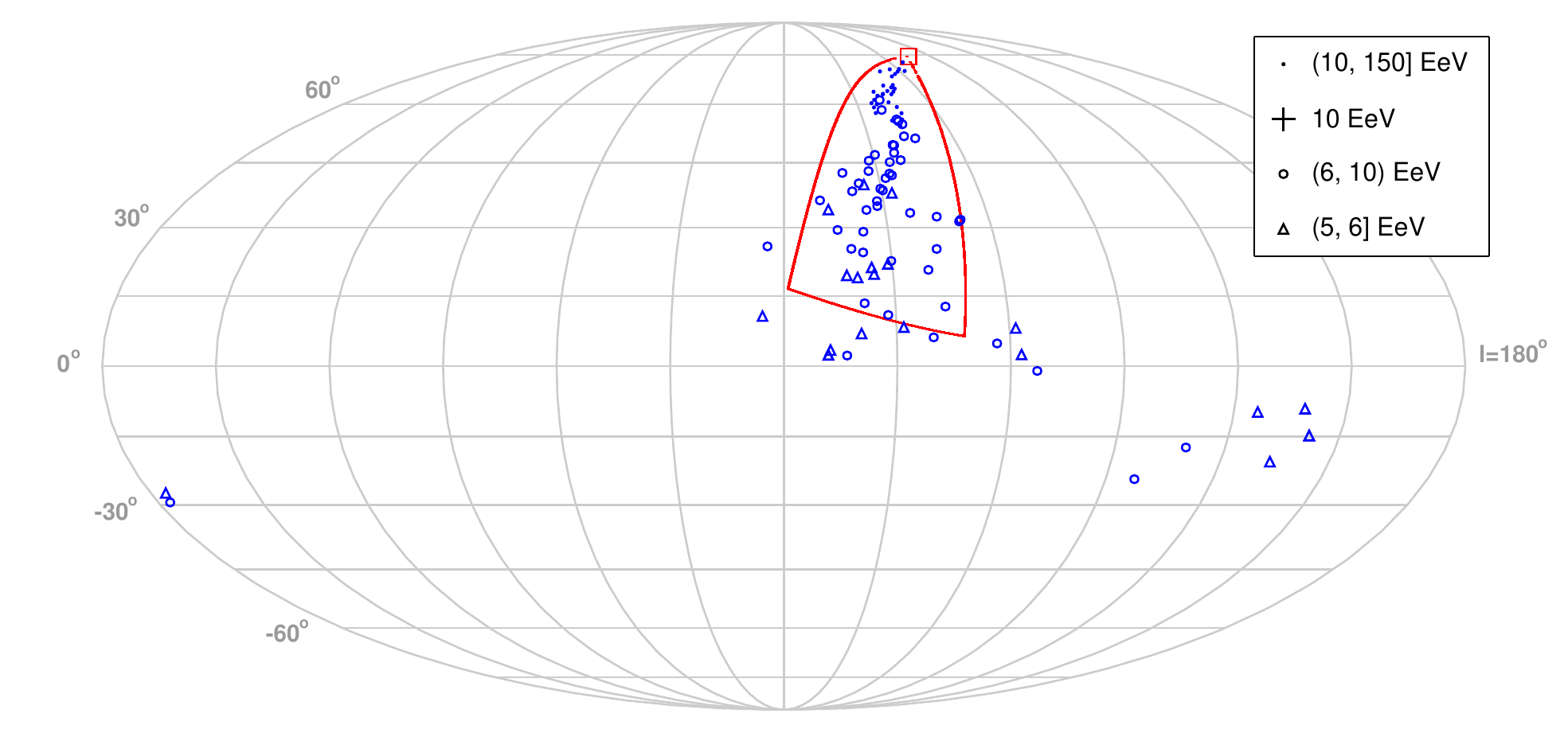}
}
\subfloat[JF12 + KRF2]{
\includegraphics[width=0.48\columnwidth]{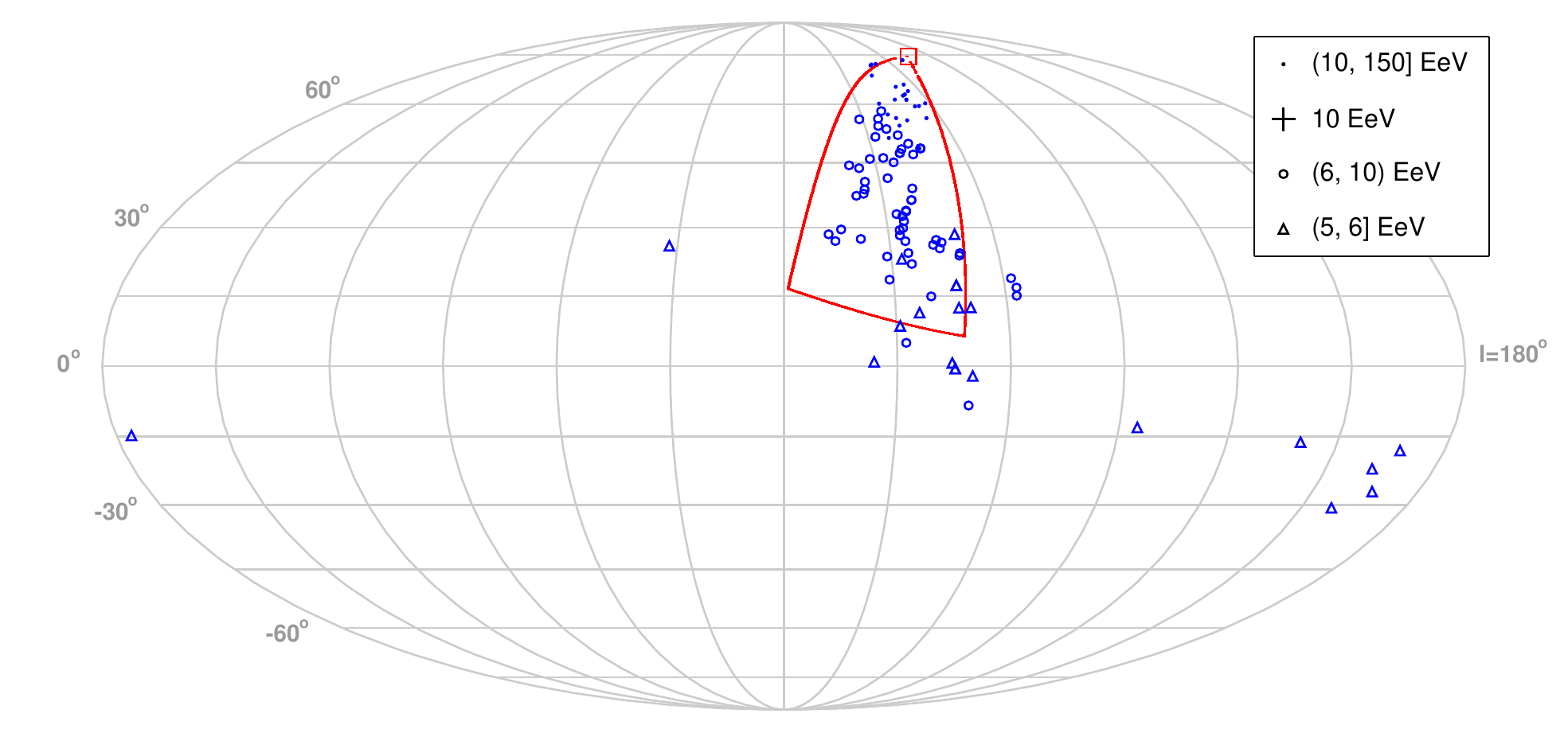}
}\\
\subfloat[JF12 + KRF3]{
\includegraphics[width=0.48\columnwidth]{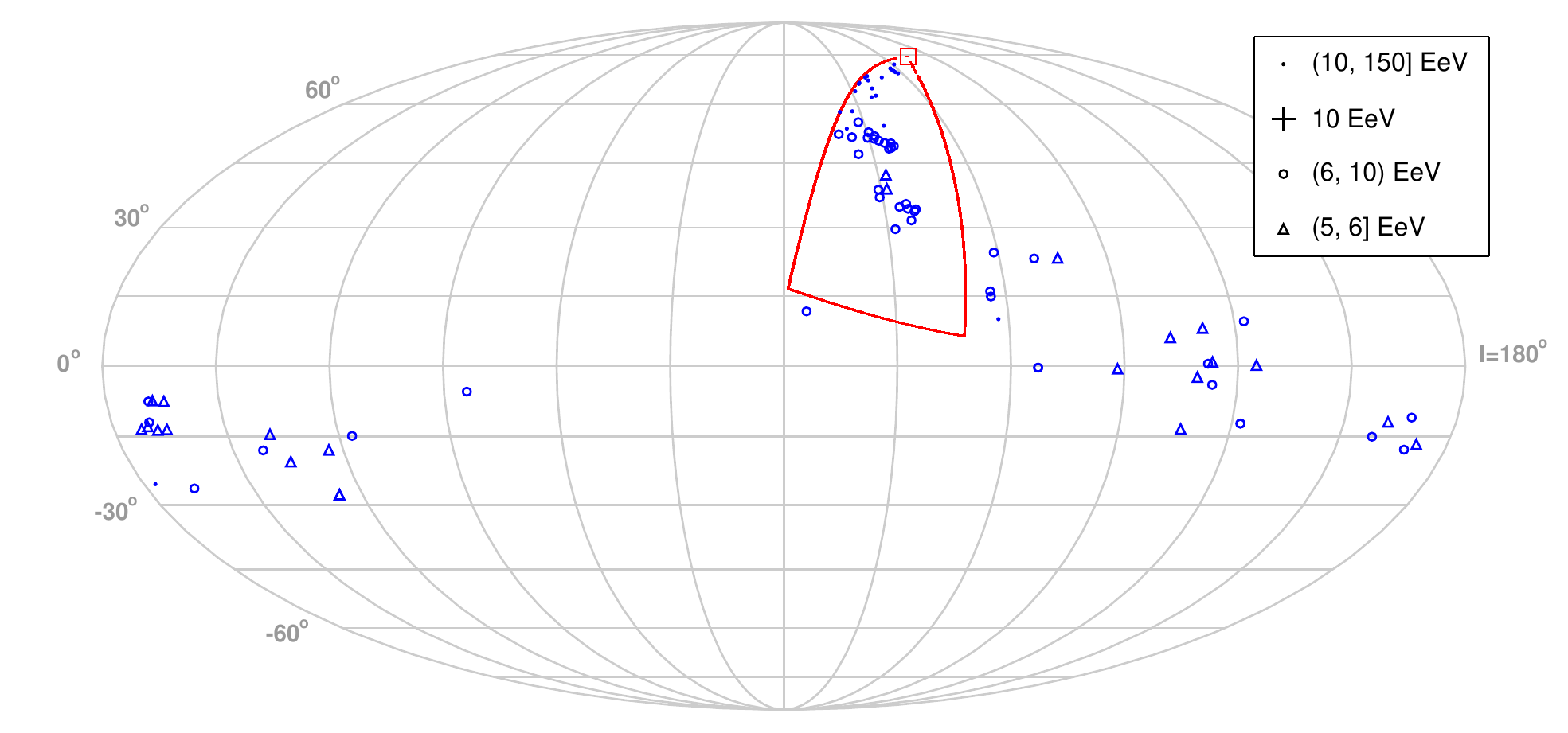}
}
\subfloat[JF12 + KRF4]{
\includegraphics[width=0.48\columnwidth]{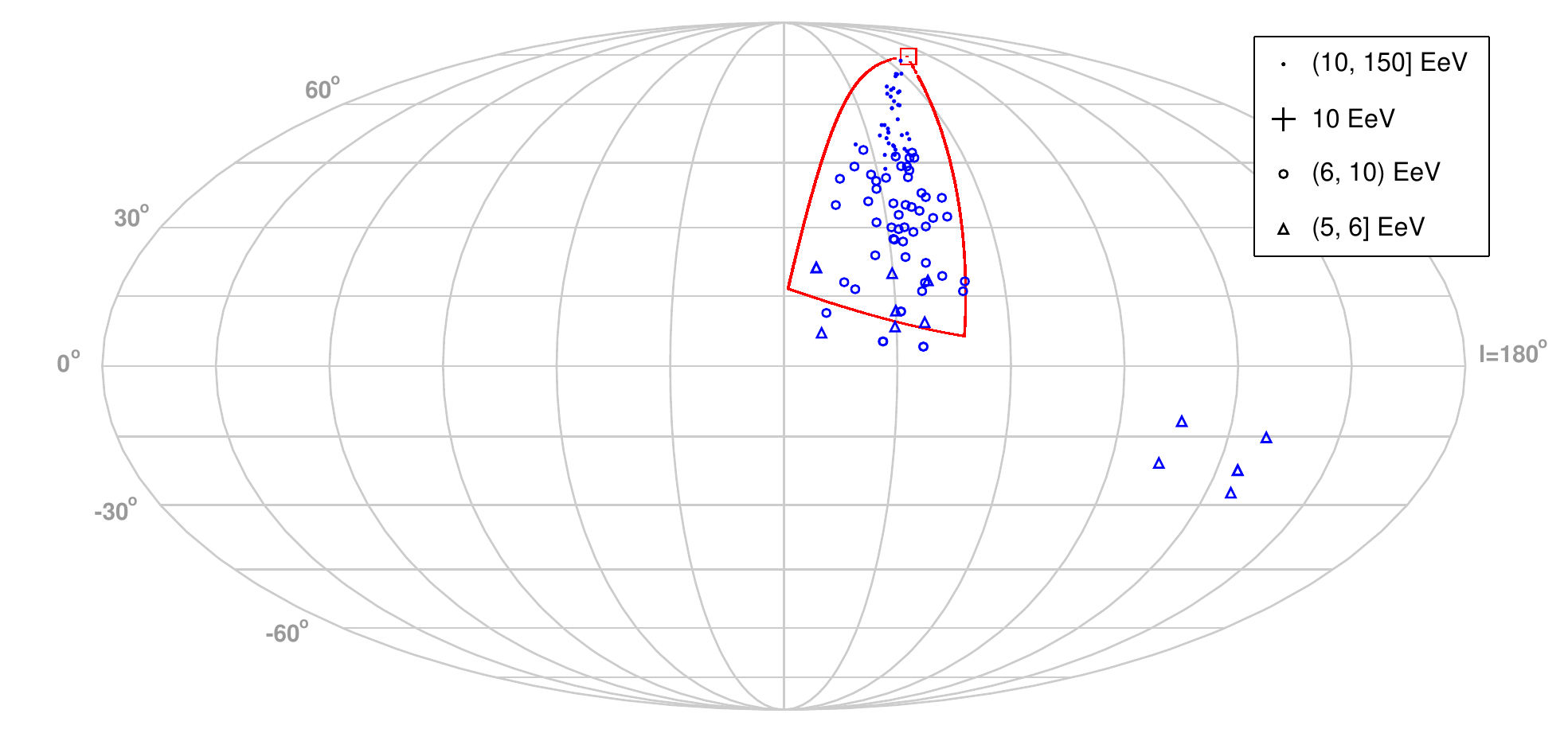}
}\\
\caption{Arrival directions for a combination of the JF12 large-scale regular
GMF field and four random field realisations. The triangular area is also shown.}
\label{Fig-map-krf}
\end{center}
\end{figure}

Arrival directions for four different realisations of the random magnetic field
are shown in Fig.~\ref{Fig-map-krf}. Our simulations show that arrival directions of
the majority of events form a triangular rather area in all four cases.
Similar observation has been presented in~\cite{Giacinti}.

This can be understood by looking in two types of deflection experienced
by UHECRs. First, a rigidity-dependent offset from the source position can be
expected due to deflections in large-scale magnetic fields. Second, a scattering
of the arrival directions caused by turbulent magnetic fields with the amplitude
inversely depending on the rigidity is expected for UHECRs from the same source.
The combination of these two effects then leads to a triangular rather than circular
image in the sky.

\begin{figure}[t!]
\begin{center}
\includegraphics[width=0.4\columnwidth]{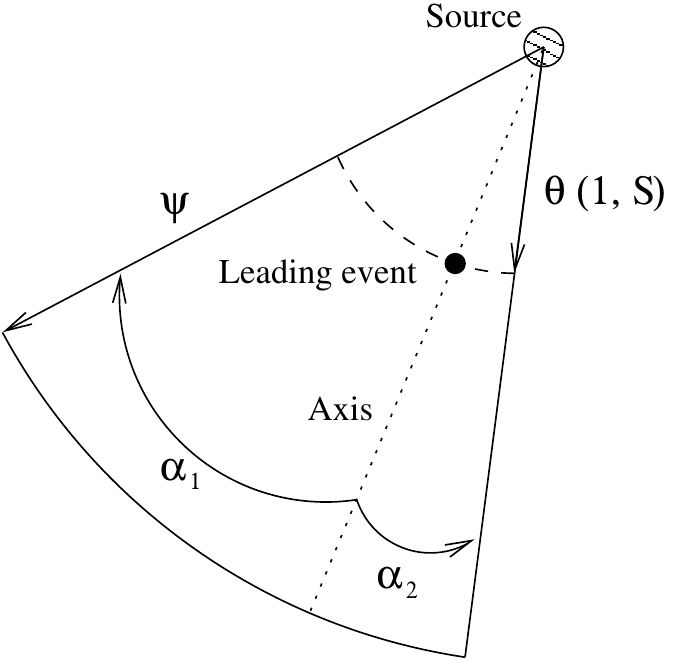}
\caption{
Sketch of the triangular area described by angles $\psi$, $\alpha_1$
and $\alpha_2$. The point source and leading event are separated
by the angular distance $\theta(1,S)$.
}
\label{Fig-scatchofarea}
\end{center}
\end{figure}

We suggest to construct the triangular area from sky positions of a source candidate
and one event. These two positions form an axis of the triangular area. The event should
lie at angular distance $\theta(1,S)$ from the source candidate larger than an angular
resolution of a cosmic-ray observation and an angular deflection expected in the GMF.
from the studied source candidate. We will call such event the leading event.
The triangular area is then defined by the axis, the maximum angular distance $\psi$
from the source candidate\footnote{The value of $\psi$ must be less than $90\degree$
to avoid shrinking of the spherical area at larger angular distances.} and two
half-opening angles $\alpha_1$ and $\alpha_2$ along the axis, see Fig.~\ref{Fig-scatchofarea}.

In Fig.~\ref{Fig-map-krf} we show the triangular area for four KRF realisations.
The same parameters were used for the triangular area in all four cases to allow
a comparison of arrival directions between our simulations. The parameters
are as follows: the leading event is $10$~EV event from the simulation for
the JF model with only regular magnetic field, $\psi=70\degree$,
$\alpha_1=\alpha_2=25\degree$.

\section{Conclusions}

Angular deflections in the GMF for the $10$ brightest radiogalaxies located within
the distance of $50$~Mpc were studied. From them Virgo~A was investigated more
extensively and even an effect of various turbulent magnetic fields was included.

We have shown that the expected UHECR image of Virgo~A has an asymmetric shape for
the magnetic rigidity above $5$~EV. If the UHECR flux is a mixture of nuclei
(i.e. protons and heavy nuclei) the image takes a triangular rather than circular
shape. This unique feature is due to high galactic latitude of Virgo~A and
can be used for an identification of Virgo~A as the UHECR point source.

Our analysis is based on the JF12 model and even if it is the most elaborated
model of the GMF, it does not provide complete picture of magnetic fields neither
in the interstellar space nor in the Galactic halo. Any discrepancy between this
GMF model and real environment may affect the arrival directions, particularly at
low rigidities. Nevertheless, our results can be adopted to any GMF model
and cosmic-ray data.

\section{Acknowledgement}

It is a pleasure to acknowledge discussions with our colleagues of the Pierre Auger
Collaboration.

\end{document}